\documentclass{aa}  

\usepackage{graphicx}
\usepackage{txfonts}
\begin{document} 

\title{Multi-frequency angular power spectrum of the 21~cm signal from the Epoch of Reionisation using the Murchison Widefield Array}
\titlerunning{Multi-frequency angular power spectrum of the 21~cm signal from the EoR using the MWA}

   \author{Cathryn M. Trott\inst{1,2}\fnmsep\thanks{Email: cathryn.trott@curtin.edu.au}
          \and
          Rajesh Mondal\inst{3,4}
          \and
          Garrelt Mellema\inst{3}
          \and
          Steven G. Murray\inst{5}
          \and
          Bradley Greig\inst{6,2}
          \and
          Jack L. B. Line\inst{1,2}
          \and
          Nichole Barry\inst{1,2}
          \and
          Miguel F. Morales\inst{7}
          }

   \institute{International Centre for Radio Astronomy Research, Curtin University, Bentley, WA 6102, Australia
         \and
            ARC Centre of Excellence for All Sky Astrophysics in 3 Dimensions (ASTRO 3D), Bentley, WA 6102, Australia
            \and
            The Oskar Klein Centre, Department of Astronomy, Stockholm University, AlbaNova, SE-10691, Stockholm Sweden
            \and
            Department of Astrophysics, School of Physics and Astronomy, Tel Aviv University, Tel Aviv 69978, Israel
            \and
            School of Earth and Space Exploration, Arizona State University, Tempe AZ, USA
        \and
        School of Physics, The University of Melbourne, 3010 Australia
        \and
        Department of Physics, The University of Washington, Seattle WA, USA}

   \date{Received ; accepted }

 
  \abstract
  {The Multi-frequency Angular Power Spectrum (MAPS) is an alternative to spherically-averaged power spectra, and computes local fluctuations in the angular power spectrum without need for line-of-sight spectral transform.  }
   {To test different approaches to MAPS and treatment of the foreground contamination, and compare with the spherically-averaged power spectrum, and the single-frequency angular power spectrum.}
   {We apply the MAPS to 110~hours of data in $z=6.2-7.5$ obtained for the Murchison Widefield Array Epoch of Reionisation experiment to compute the statistical power of 21~cm brightness temperature fluctuations. In the presence of bright foregrounds, a filter is applied to remove large-scale modes prior to MAPS application, significantly reducing MAPS power due to systematics. }
   {The MAPS shows a contrast of 10$^2$--10$^3$ to a simulated 21~cm cosmological signal for spectral separations of 0--4~MHz after application of the filter, reflecting results for the spherically-averaged power spectrum. The single-frequency angular power spectrum is also computed. At $z=7.5$ and $l=200$, we find an angular power of 53~mK$^2$, exceeding a simulated cosmological signal power by a factor of one thousand. Residual spectral structure, inherent to the calibrated data, and not spectral leakage from large-scale modes, is the dominant source of systematic power bias. The single-frequency angular power spectrum yields slightly poorer results compared with the spherically-averaged power spectrum, having applied a spectral filter to reduce foregrounds. Exploration of other filters may improve this result, along with consideration of wider bandwidths.}
   {}


   \maketitle
%

\section{Introduction}
\label{sec:int}
Exploration of the Epoch of Reionisation (EoR; $z=5.3-10$) through the 21~cm neutral hydrogen transition remains a primary goal of many current and upcoming interferometric low-frequency radio telescopes, including the Murchison Widefield Array \citep[MWA,][]{tingay13_mwasystem,bowman13_mwascience,wayth18}, LOFAR{\footnote[1]{http://www.lofar.org}} \citep{vanhaarlem13}, Hydrogen Epoch of Reionization Array (HERA){\footnote[3]{http://reionization.org}} \citep{deboer17} and the upcoming Square Kilometre Array (SKA){\footnote[4]{http://skatelescope.org}} \citep{koopmans15}. Evolution of the signal over redshift is also explored through the global spatially-averaged temperature, using both single element and short-spacing interferometric arrays \citep{edges18,saras322,2022arXiv220310466T}. The EDGES experiment \citep{edges18} reported the detection of a deep absorption trough at 78~MHz, suggested to mark the birth of the first stars and the commencement of Lyman-$\alpha$ coupling, however the cosmological origin of this measurement has been disputed by recent measurements by the SARAS3 experiment \citep{saras322}.

These substantial international efforts have steadily moved the field closer to the detection, and characterisation, of this cosmological signal, but as yet experiments have not reported success, being hampered by the complex structure of the bright foreground signal from radio galaxies and our Galaxy, and the difficulties associated with performing precision experiments with low-frequency radio telescopes. As such, the use of even seemingly simple metrics, such as the spatial power spectrum, have not yielded success.

The Multi-Frequency Angular Power Spectrum (MAPS) was first proposed under that name by \citet{2007MNRAS.378..119D}, although it had been studied earlier by for example \citet{2005MNRAS.356.1519B} and \citet{2005ApJ...625..575S}. \citet{mondal18,mondal19} developed it further and a recent paper provided the first study of its performance in the estimation of EoR parameters using an MCMC framework \citep{mondal22}.
\citet{pal21} recently applied MAPS to 150~MHz data from the GMRT telescope, using the Tapered Gridded Estimator, and placing limits on the power spectrum at $z=8.28$. MAPS computes the angular power spectrum as a function of spectral separation, characterising the 21~cm signal correlation as a function of scale. It is a useful statistic in the presence of light-cone effects, whereby the evolution of signal along the line-of-sight destroys signal ergodicity for three-dimensional power spectrum analyses. MAPS is straightforward to implement, without need for a line-of-sight transform, which otherwise can be problematic. Application to real data, however, is complicated by spectrally-correlated continuum foreground sources, which can be 3--4 orders of magnitude brighter than the EoR signal.

In this work, we apply MAPS to 110 hours of data from the EoR0 observing field obtained with the MWA between 2013 and 2016 in frequency range 167--197~MHz ($z=6.2-7.5$). MAPS is applied before and after application of a foreground filter, which is designed to suppress structure on delays shorter than a set value (e.g., 180~ns), while retaining signal on faster varying scales.

The MAPS software is available from Github\footnote{\url{https://github.com/rajeshmondal18/MAPS}}, however this uses image-based datacubes as inputs. We will apply the MAPS algorithm to gridded visibility data from the MWA experiment, and therefore develop our own software, following the algorithm described in \citet{mondal20,mondal22}. The dataset is identical to that used in \citet{trott20} to extract cylindrically- and spherically-averaged Fourier power spectra.

\section{METHODS}
The single-frequency, $\nu$, angular power spectrum is defined by:
\begin{equation}
    C_l^s(\nu) = C_{2\pi U}(\nu) = \frac{1}{\Omega}\langle \tilde{T}_b({\bf U})\tilde{T}_b(-{\bf U}) \rangle,
\end{equation}
for beam field-of view $\Omega$ sr and angular multipole $l=2\pi U$.
By extension, MAPS is defined by \citet{mondal18}:
\begin{equation}
    C_l(\nu_1,\nu_2) = C_{2\pi U}(\nu_1,\nu_2) = \frac{1}{\Omega}\langle \tilde{T}_b({\bf U},\nu_1)\tilde{T}_b(-{\bf U},\nu_2) \rangle.
\end{equation}
If we impose ergodicity and periodicity along the frequency direction, we have $C_l(\nu_1, \nu_2) \equiv C_l(\Delta \nu)$. The dimensionless MAPS is:
\begin{equation}
    \mathcal{D}_l(\nu_1, \nu_2) = \frac{l(l+1)C_l(\nu_1, \nu_2)}{2\pi},
\end{equation}
which has units of temperature squared. Similarly, the dimensionless single-frequency angular power spectrum is computed likewise:
\begin{equation}
        \mathcal{D}_l^s(\nu) = \frac{l(l+1)C_l^s(\nu)}{2\pi}.
        \label{eqn:aps}
\end{equation}

With an interferometer, we measure visibilities (coherence of the electric field) as a function of scale (baseline length), in units of Jansky. For power spectrum estimation, it is typical to grid measurements onto a common $uv$-plane to allow for coherent addition of data (lower noise). For this, a gridding kernel, $K$, is employed, which matches or mimics the Fourier representation of the instrument primary beam\footnote{In an optimal power spectrum estimator, the gridding kernel is the Fourier Transform of the telescope response function to the sky (i.e., the beam), because this correctly represents the smearing of information due to the finite station size. In practise, the MWA beam is highly-structured with sidelobes, and we instead employ a size-matched Blackman Harris 2D window function as the kernel, as is done in \citet{trott16chips}.}. The averaged sky signal after this process for cell $u,v$ is given by:
\begin{equation}
    \mathcal{V}(u,v) = \frac{\sum_i V(u_i,v_i)K(u-u_i,v-v_i)}{\sum_i K(u-u_i,v-v_i)} \,\,\text{Jy}.
\end{equation}
In order to remove power bias due to additive noise, power spectrum estimators, such as CHIPS \citep{trott16chips} also typically use some spectral or temporal differencing such that data with different noise realisations, but matched signal, are multiplied,
\begin{equation}
    P(k) \propto \langle \mathcal{V}(t_1) \mathcal{V}^\ast(t_2) \rangle = \frac{1}{4}(P_{tot}(k) - P_{diff}(k)),
\end{equation}
where $P_{tot} = |\mathcal{V}(t_1) + \mathcal{V}^\ast(t_2)|^2$ and $P_{diff} = |\mathcal{V}(t_1) - \mathcal{V}^\ast(t_2)|^2$, which can be numerically easier to implement. CHIPS outputs these gridded total and differenced visibilities, and their weights.

For gridded visibility data, the MAPS algorithm can be applied directly, such that:
\begin{equation}
    C_l(\nu_1,\nu_2) = \frac{1}{\Omega}\frac{\sum \mathcal{V}(U,\nu_1)\mathcal{V}^\ast(-U,\nu_2)W(\nu_1)W(\nu_2)}{\sum W(\nu_1)W(\nu_2)},
\end{equation}
where $W(\nu)$ gives the weight at that frequency.
For time-interleaved data, the final power is:
\begin{equation}
    C_l(\nu_1,\nu_2) = \frac{1}{4}(C_{l,tot} - C_{l,diff}),
\end{equation}
and the dimensionless form is:
\begin{equation}
    \mathcal{D}_l(\nu_1,\nu_2) = \frac{l(l+1)}{8\pi}(C_{l,tot}(\nu_1,\nu_2)-C_{l,diff}(\nu_1,\nu_2)) \,\,\,\, \text{mK}^2.
    \label{eqn:dcl}
\end{equation}
Equation \ref{eqn:dcl} computes the local MAPS between all sets of spectral channels, providing the greatest flexibility to studying the signal, and natively allowing for signal evolution, but at the expense of signal-to-noise. I.e., each spectral channel difference is individually computed, and therefore the evolution of a given $\Delta\nu$ can be studied. Alternatively, one may consider a smaller cube, where the signal is ergodic throughout (i.e., no significant signal evolution from the back to the front of the cube), and spectral differences stacked to increase signal-to-noise ratio. In this calculation, any evolution of signal on a given scale is erased. This latter approach matches more closely the conditions for a spherically-averaged power spectrum analysis, as shown by \citet{mondal22}. In this case, we compute:
\begin{equation}
    \mathcal{D}_l(\Delta\nu) = \frac{l(l+1)\displaystyle\sum_{\nu_1,\nu_2}(C_{l,tot}(\Delta\nu)-C_{l,diff}(\Delta\nu))}{8\pi} \,\,\,\, \text{mK}^2.
    \label{eqn:ergodic}
\end{equation}
Both approaches are implemented in this work. The noise is obtained by considering the differenced visibilities:
\begin{equation}
    \Delta \mathcal{D}_l(\nu_1,\nu_2) = \frac{l(l+1)}{8\pi}(C_{l,diff}(\nu_1,\nu_2)) \,\,\,\, \text{mK}^2.
\end{equation}

\subsection{Foregrounds}
As discussed by \citet{mondal22}, foreground contaminating sources are problematic for MAPS due to their large spectral coherence (continuum sources). Despite some bright point sources being removed from the dataset, foregrounds remain a significant problem for 21~cm studies. To combat this, we employ a non-parametric foreground that has been designed to match that for band-limited discrete flat spectrum sources; DAYENU. The DAYENU filter \citep{dayenu} was introduced to provide a clean filter that suppresses signal on line-of-sight modes slower than a defined delay, while leaving faster modes (including those with EoR signal) unfiltered. We employ an adjusted version of this filter to suppress foregrounds. Missing frequency channels are first handled through a least squares \citep[CLEAN; ][]{parsons09} solution, such that we apply a restoration filter:
\begin{equation}
    \mathcal{R} = \mathcal{I}w - \mathcal{R}_D,
\end{equation}
where,
\begin{equation}
    \mathcal{R}_D = A (A^T w A)^{-1} A^T w \mathcal{I},
\end{equation}
where $A$ is a matrix of the eigenvectors of the DAYENU matrix (Discrete Prolate Spheroidal Series),
\begin{equation}
    \mathcal{C}_{mn} = \frac{2\pi\Delta\nu}{\epsilon} \mathrm{sinc}(2\pi\tau(\nu_m-\nu_n)),
\end{equation}
which operates on the spectral data. Fundamentally, the sinc function is the Fourier Transform of the Heaviside function of a set of foregrounds extending to a fixed delay (e.g. the horizon) in the power spectrum. The delay ($\tau$ ns) and depth ($\epsilon$) of the filter can be adjusted to shape the filter. A value of $\epsilon=10^9$ is used throughout.

\subsection{Data}
The data used for this work are those processed to a Fourier power spectrum in \citet{trott20}. These data were obtained with the MWA over 2013--2016, and comprise 110~hours of observation on the EoR0 field (RA=0h, Dec=$-$27deg) spanning 30.72~MHz from 167--197~MHz, with 80~kHz spectral resolution. Only the East-West polarization is used, because it has much lower response to power from the Galaxy for this field. MWA data contain regular missing spectral channels due to the channelisation filter, yielding two missing channels each 1.28~MHz bandwidth. These are treated as part of the filter process, but are omitted in calculation of the MAPS (assigned zero weights). These data were calibrated with the RTS software pipeline, with 1000 sources subtracted (residual source flux density less than 50~mJy). 

In addition to the data, we also have a wide ($\sim7.5$~Gpc on a side) 21cmFAST \citep{mesinger11} simulation cube designed to match the large MWA primary beam and bandwidth in the 167--197~MHz frequency range \citep{greig22}. 21cmFAST efficiently generates 21~cm brightness temperature cubes using an excursion set formalism to calculate ionisations by UV photons emitted from galaxies described using a simple, six parameter astrophysical model. Specifically, these describe both a mass-dependent star-formation and escape fraction efficiency, a star-formation time-scale and a minimum mass threshold for active star-forming galaxies. Here we use the parameters defined in \citet{barry19} consistent with those from \citet{park19}. This simulation additionally assumes that the intergalactic neutral gas is sufficiently heated by a background heating source (e.g. X-rays) such that the spin temperature is larger than the cosmic-microwave background. The 21cmFAST cube has been projected onto 384 Healpix nside 2048 maps, each corresponding to a frequency channel matching those of the real data detailed above. These maps were fed into the \texttt{WODEN} simulator~\citep{Line2022}, with each pixel in the Healpix maps input as a point-source (a delta function). \texttt{WODEN} then calculates the measurement equation for all directions, essentially performing a direct Fourier transform for $\sim$ 25 million directions per frequency channel. In addition, a frequency-interpolated version of the MWA Fully Embedded Element primary beam model was calculated for all directions and applied to add the instrumental response. The simulated outputs can be processed through the same framework, using the gridded visibilities to produce the MAPS.

\section{RESULTS}
\label{sec:res}
\subsection{Filter}
\begin{figure}[hbt!]
\centering
\includegraphics[width=0.48\textwidth]{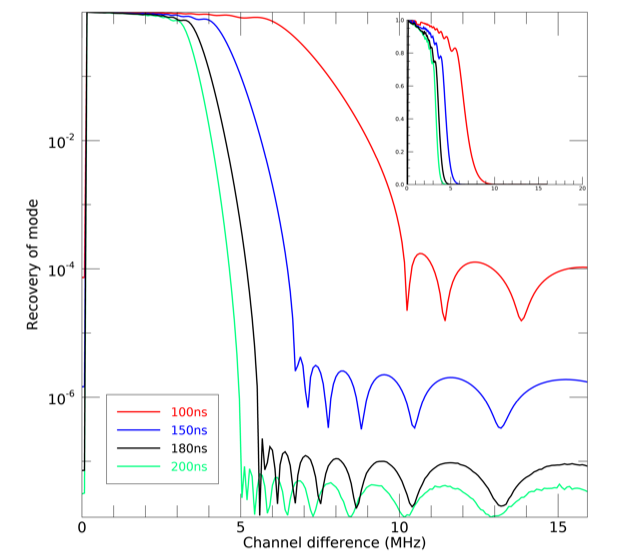}
\caption{Recovery of signal power in a given spectral differencing mode after application of a foreground filter to the full-band data (30.72~MHz). Inset shows the same data with a linear scaling. A 180~ns delay filter provides sufficient recovery on scales of interest ($\Delta\nu <$4~MHz), but with a small correction required.}
\label{pdur}
\end{figure}

The filter is designed to suppress structure on scales with delay smaller than a specified value. For this work, we aim to suppress the large-scale modes, while retaining signal on spectral separations smaller than $\sim$5~MHz; \citet{mondal22} shows that little power is expected on larger separations. The filter is applied to the full-band (30.72~MHz) data, prior to any sub-division of the band to smaller cubes (8--10~MHz). This produces the cleanest filter response. Figure~\ref{pdur} shows the recovery of MAPS as a function of spectral separation, for four different delays that are typical for foreground-dominated modes - 200~ns, 180~ns, 150~ns, 100~ns. The recovery is computed as the ratio of the post-filter to the pre-filter power in a unity-amplitude complex sinusoid, $s(\nu,\tau)=\exp{2\pi{i}\nu\tau}$.
The 180~ns delay is chosen herein because it has excellent recovery below 5~MHz, but with the maximum suppression of larger scales. Despite the excellent recovery, there is some signal loss, which is corrected after application of the filter.

We start by demonstrating the effect of the filter on the output of the MAPS algorithm. The ergodic MAPS algorithm is considered, most akin to the power spectrum, whereby only the spectral difference is considered, and the data are aggregated. 

\subsection{MAPS - no filter}
\begin{figure*}[hbt!]
\centering
\includegraphics[width=0.9\textwidth]{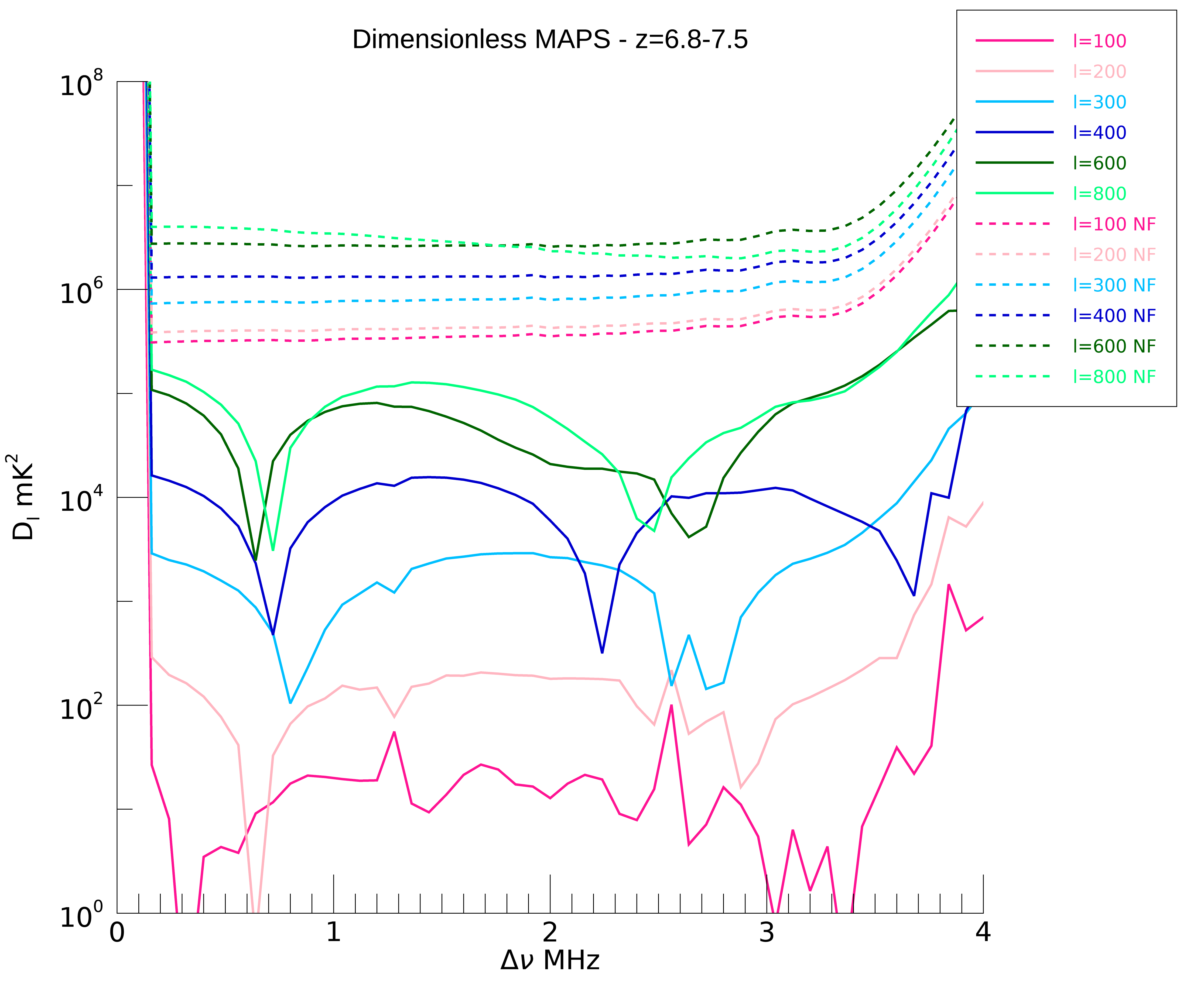}
\caption{Comparison of MAPS outputs without (dashed) and with (solid) the 180~ns filter for 192 channels in $z=6.8-7.5$ for eight measured $l$-bins, where `NF' denotes no filter being applied.}
\label{fig:filtercompare}
\end{figure*}

Figure \ref{fig:filtercompare} (dashed) shows the output of MAPS for the same 192 channels (15~MHz) at the lower end of the high-band data ($z=6.8-7.5$), but without the 180~ns filter, for eight computed $l$ modes. Note that $l=100$ corresponds to $k_\bot=l/DM=0.017 h$~Mpc~$^{-1}$.
The power is consistently high across all estimated $l$-modes, with performance worsening on scales larger than 3~MHz. These values exceed the expected 21~power by at least six orders of magnitude, and are not useable for science.

\subsection{Filtered MAPS -- ergodic}
The output of Equation \ref{eqn:ergodic} is plotted for the first eight $l$ modes for the same $z=6.8-7.5$ set with the 180~ns filter, using logarithmic scales (Figure \ref{fig:filtercompare}, solid lines). The absolute value is plotted, but, in general, some modes are negative owing to the filter response. A correction is applied to the power for spectral differences smaller than 4~MHz to alleviate the small signal loss that occurs due to the filter. It is clear that application of the filter reduces the signal by approximately 2--5 orders of magnitude depending on scale, and therefore is worthwhile to apply. There are noticeable features present in the filtered MAPS, however, and these correspond to the oscillatory structure observed in the filter response in Figure \ref{pdur}. Further tuning of the filter may be of benefit, but for these data is unlikely to yield further improvements.

\begin{figure}
\centering
\includegraphics[width=0.48\textwidth]{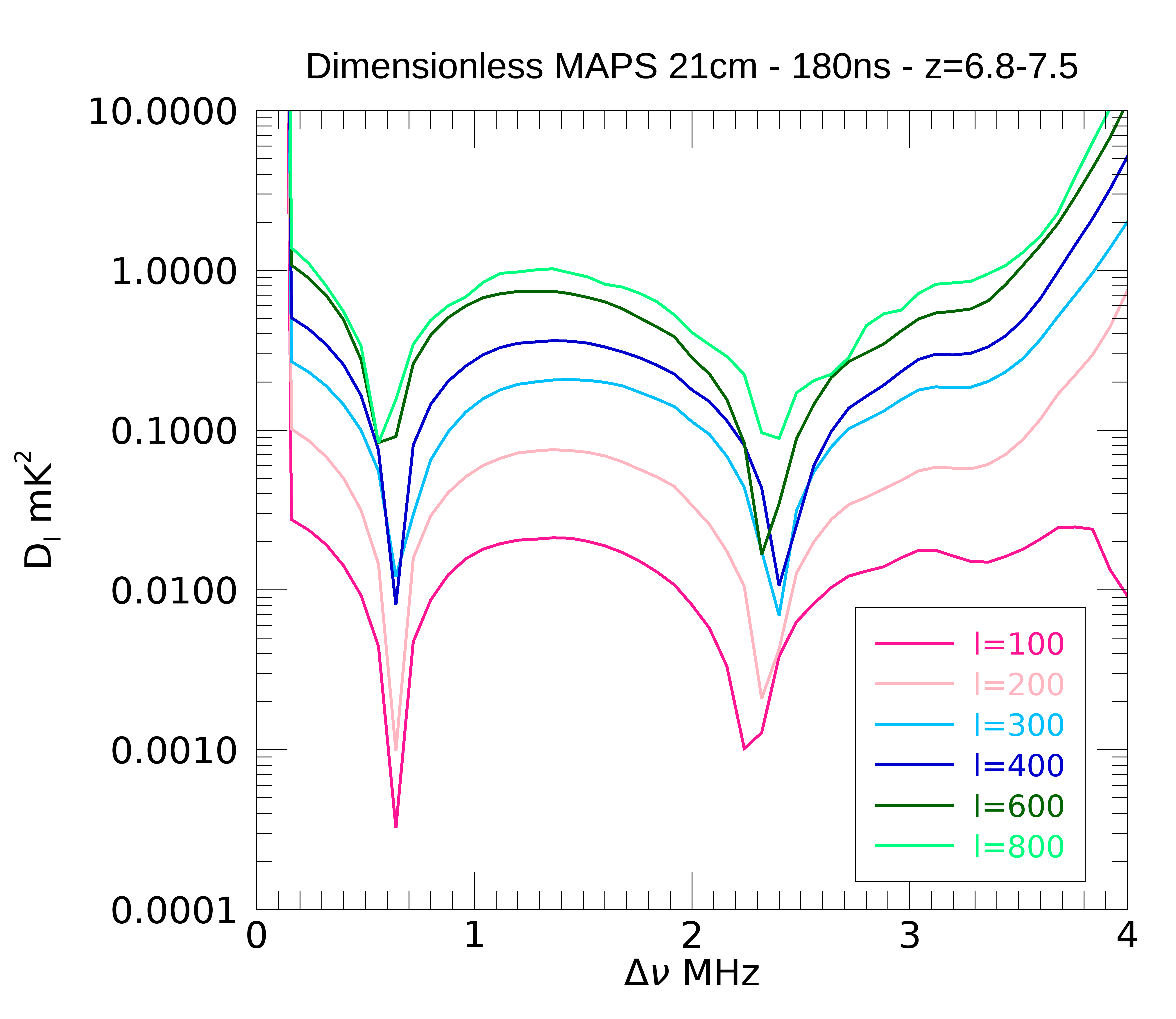}
\caption{Comparison 21cmFAST simulated MAPS for $z=6.8-7.5$.}
\label{fig:21cm}
\end{figure}

Excellent results are produced for $l=100, 200$, with larger modes showing commensurately poorer results in line with the $l^2$ scaling of the dimensionless MAPS. As a comparison, the 21cmFAST cube output is shown for the same redshift range, and binning (Figure \ref{fig:21cm}).
The results are similar to those in \citet{mondal22}, but with lower amplitude, commensurate with the higher redshift of this cube. In general, the data yield MAPS amplitudes two orders of magnitude higher than the simulated signal, matching the Fourier power spectrum results produced for the \citet{trott20} work.

\begin{figure*}
\centering
\includegraphics[width=0.48\textwidth]{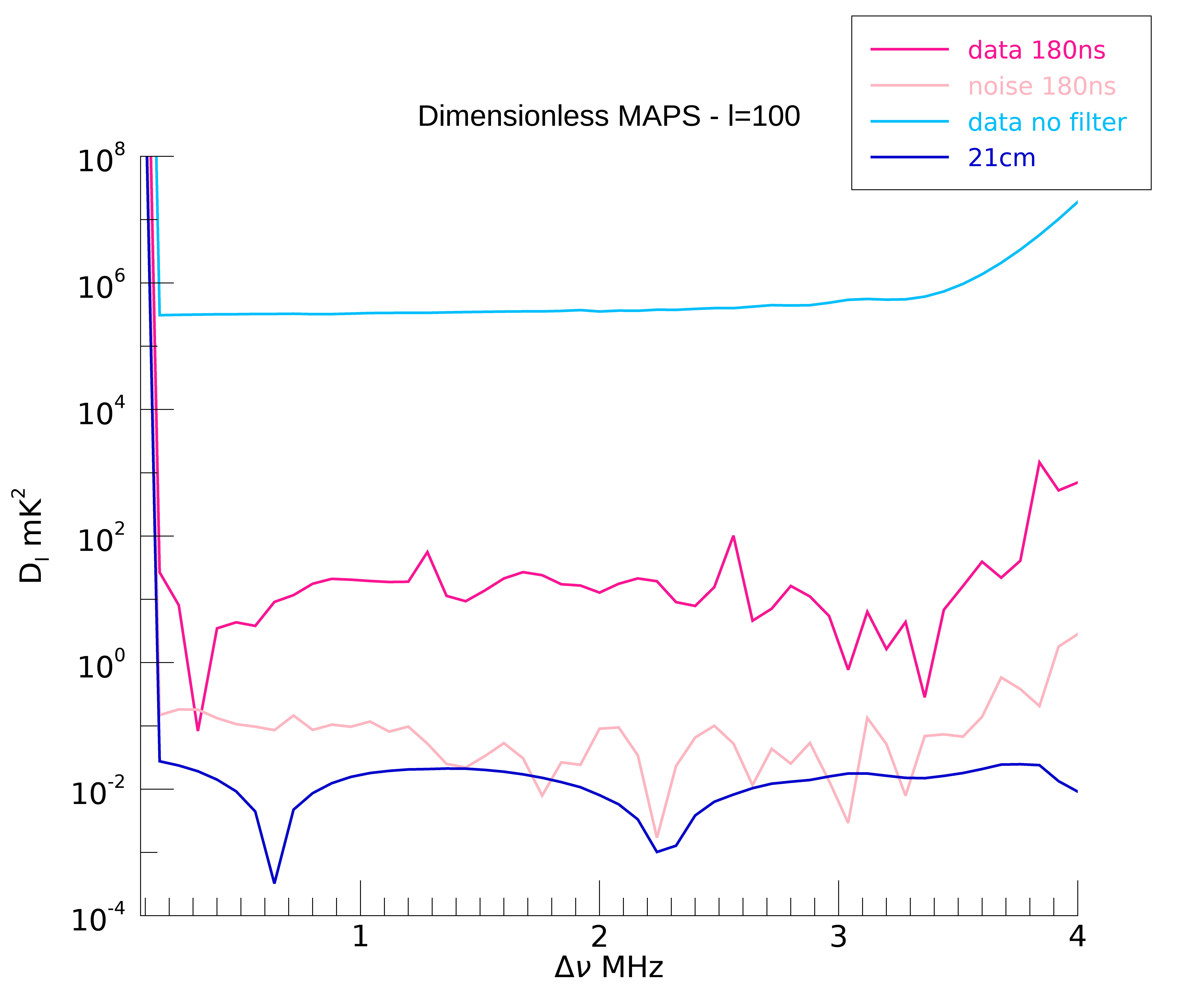}
\includegraphics[width=0.48\textwidth]{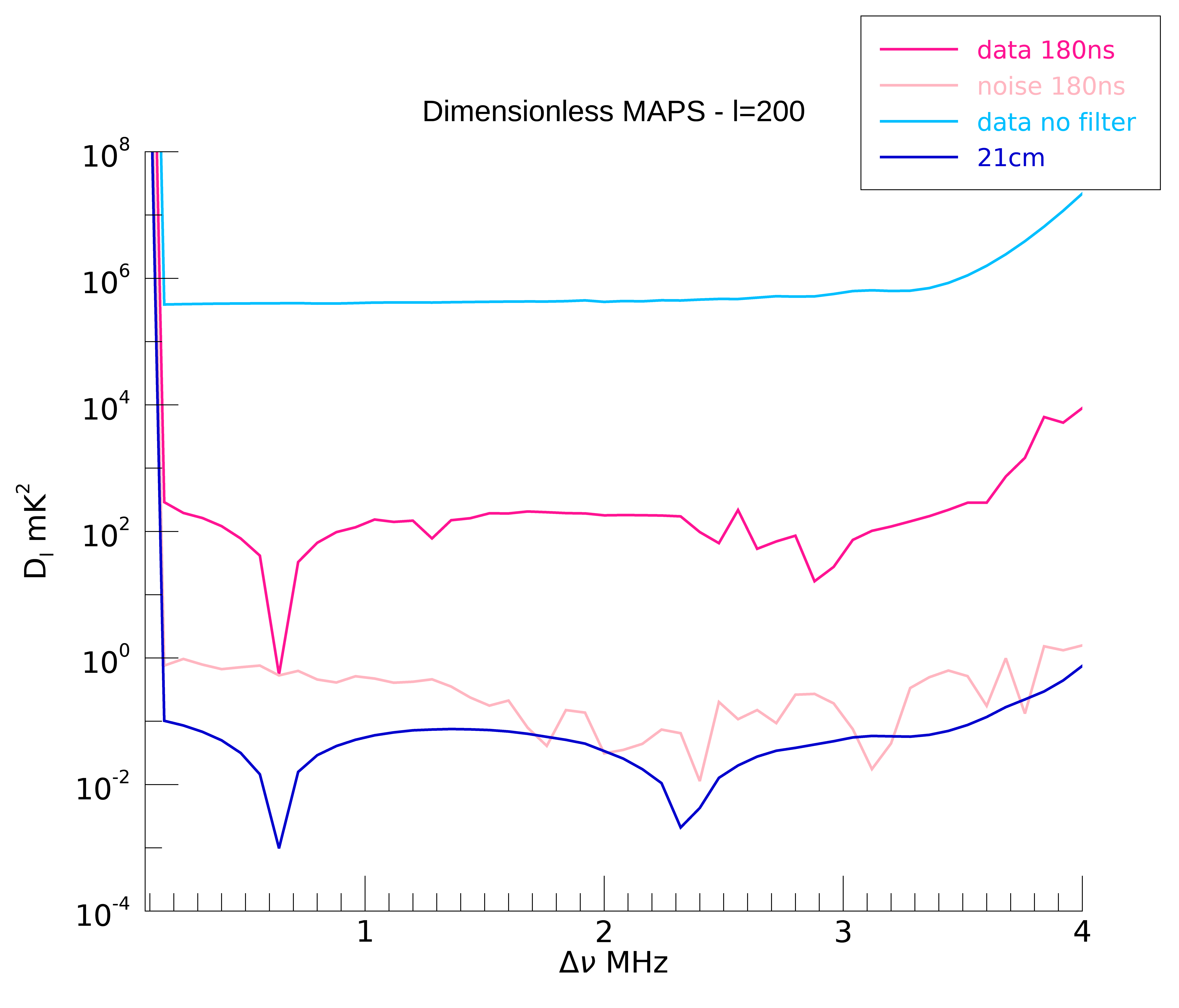}\\
\includegraphics[width=0.48\textwidth]{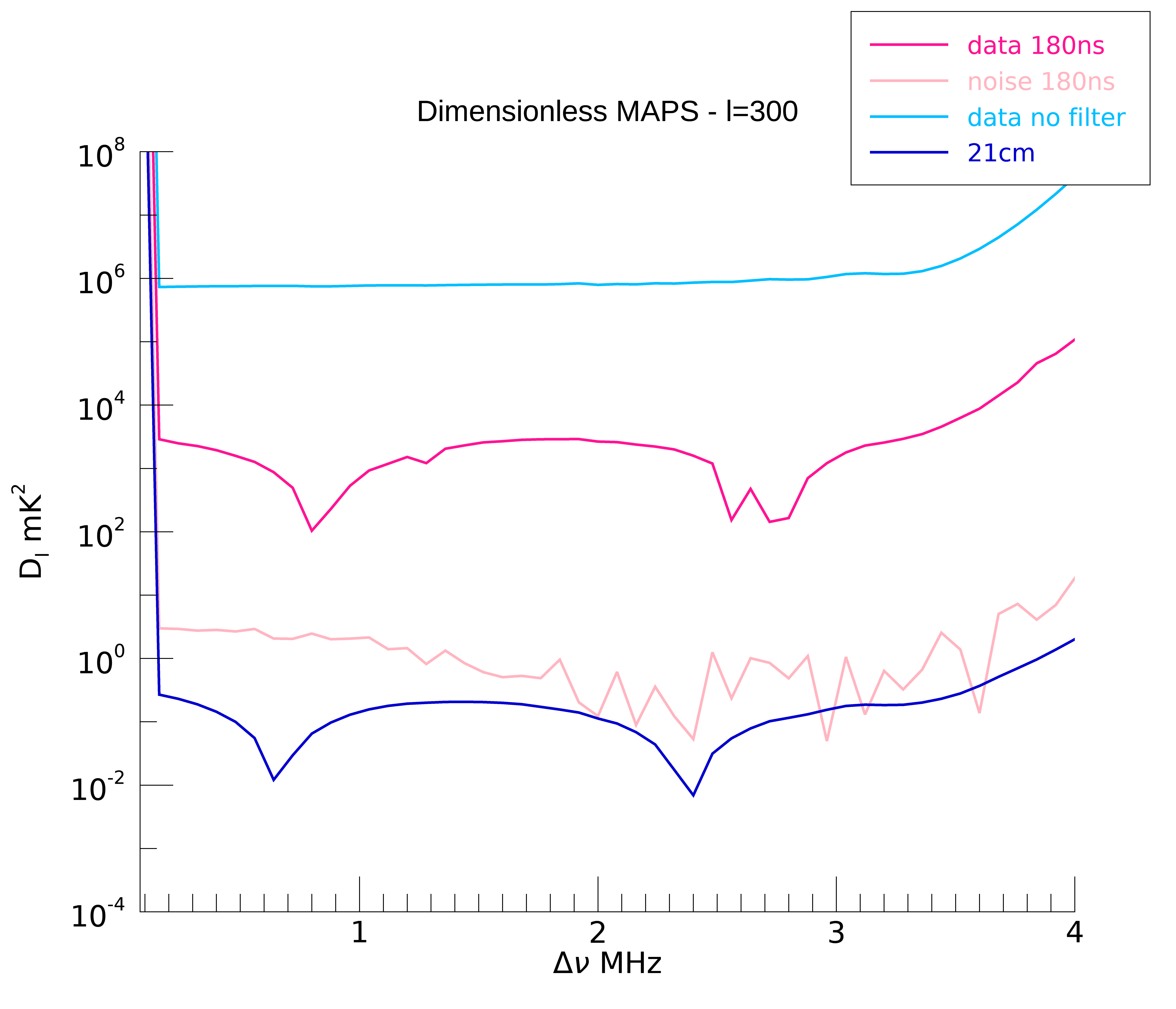}
\caption{Dimensionless MAPS for $z=6.8-7.5$ for 192 80kHz channels and the first three $l$-modes, showing the full data (no filter), full data (180~ns filter), the 21cm simulation (with filter), and the data noise level.}
\label{fig:lrange}
\end{figure*}

We also plot the first three $l$-modes for the full data (no filter), full data (180~ns filter), the 21cm simulation (with filter), and the data noise level, in Figure \ref{fig:lrange}.
The filter is again shown to have an impact on the contaminating power, but the data remain above the expected noise level, and the 21~cm signal power. The measured power bias relative to the thermal noise is evidence of residual systematics in the data. These plots also demonstrate that these data are not sufficient to detect this model signal, even in the absence of systematics, but are within an order of magnitude.

\subsection{Filtered MAPS -- local}
The local MAPS treats the power for each set of two spectral channels individually, without averaging. Without the assumption of ergodicity, it can be applied to the full-band (30.72~MHz) data as long as we only consider spectral differences for which the filter does not suppress signal ($\Delta\nu <$ 4~MHz).

\begin{figure*}
\centering
\includegraphics[width=0.48\textwidth]{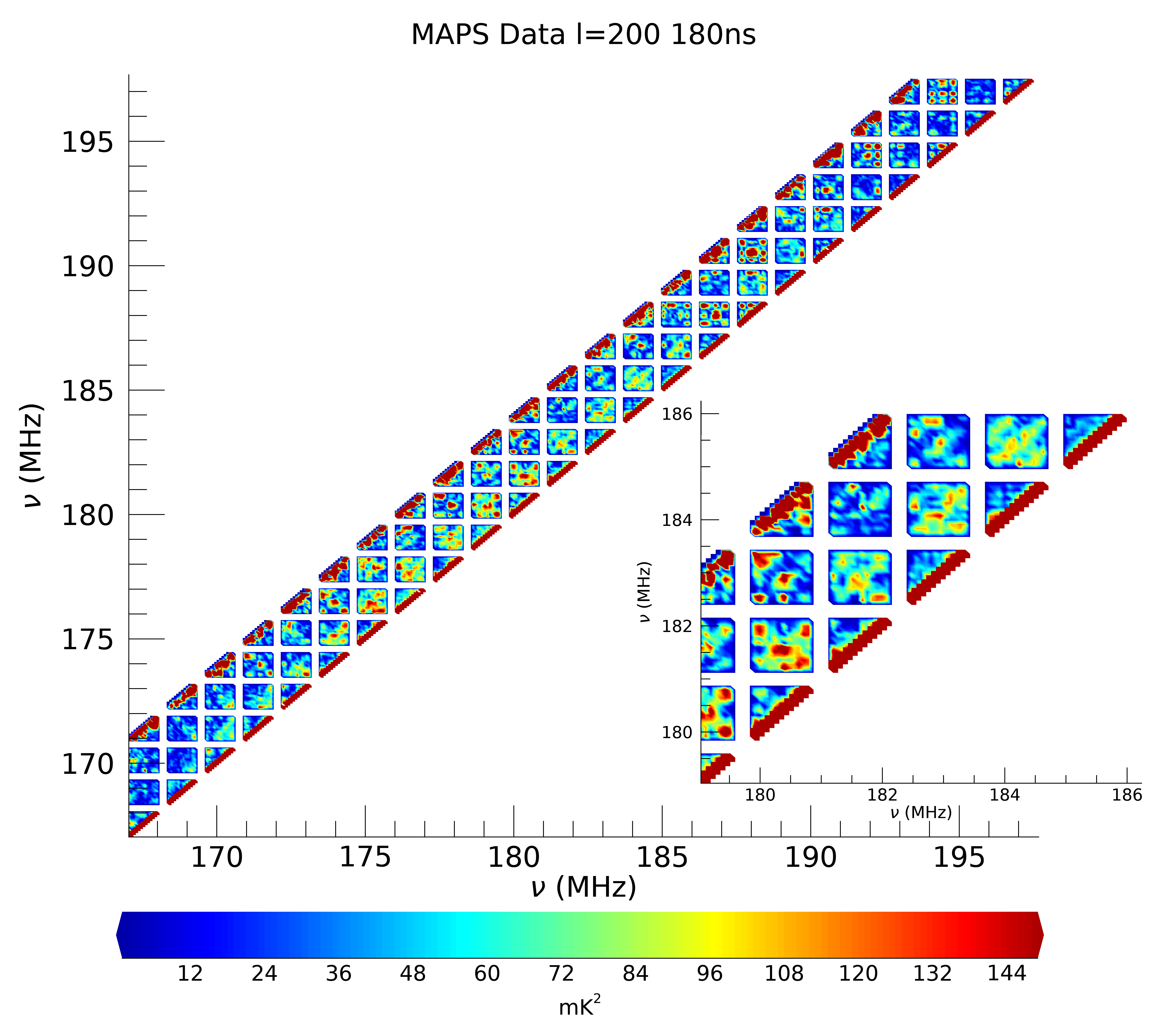}
\includegraphics[width=0.48\textwidth]{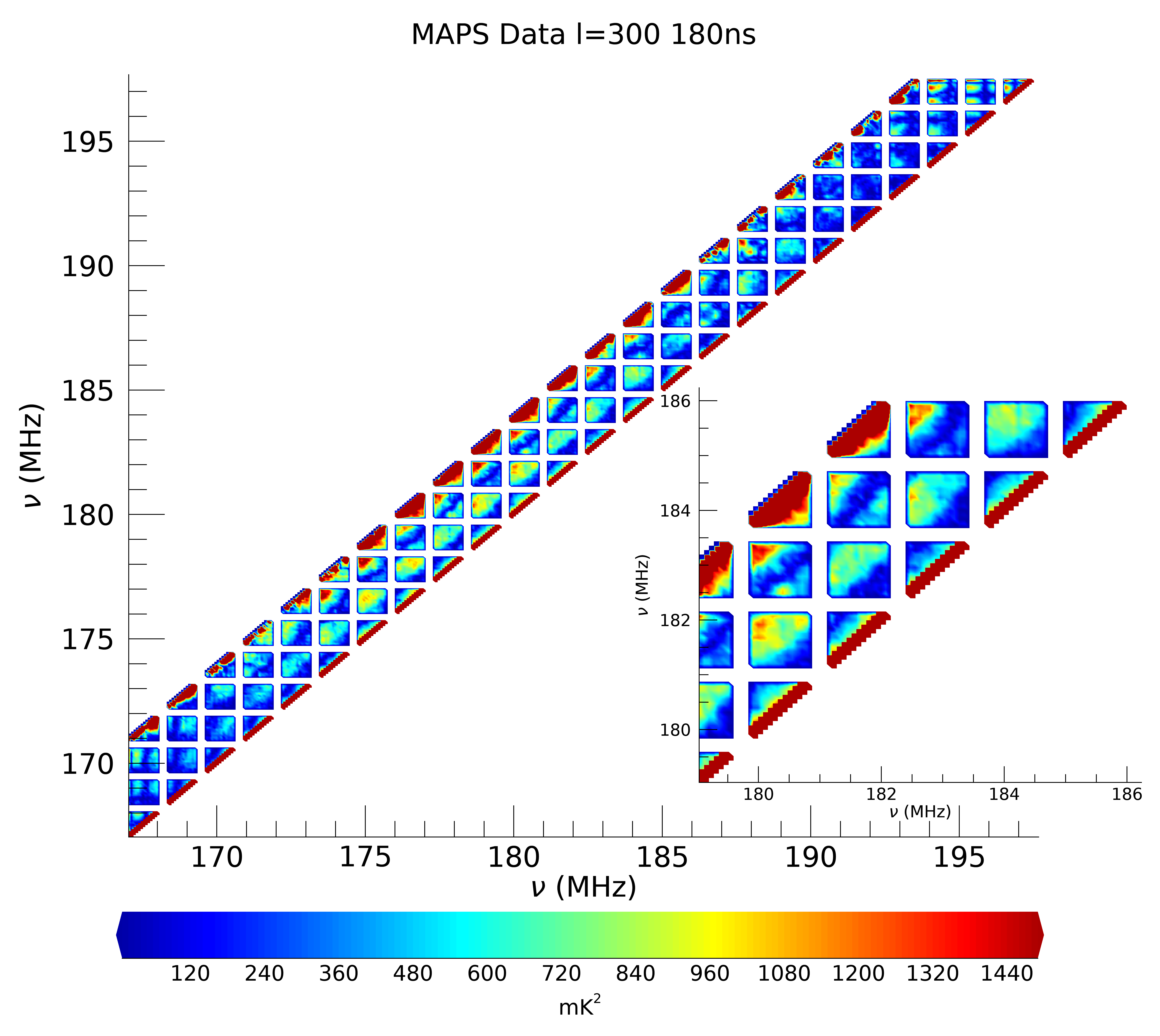}
\caption{Contour maps of dimensionless MAPS for the data and $l=$200, 300. Note the different colour scale for each plot.}
\label{fig:2d_data}
\end{figure*}

\begin{figure*}
\centering
\includegraphics[width=0.48\textwidth]{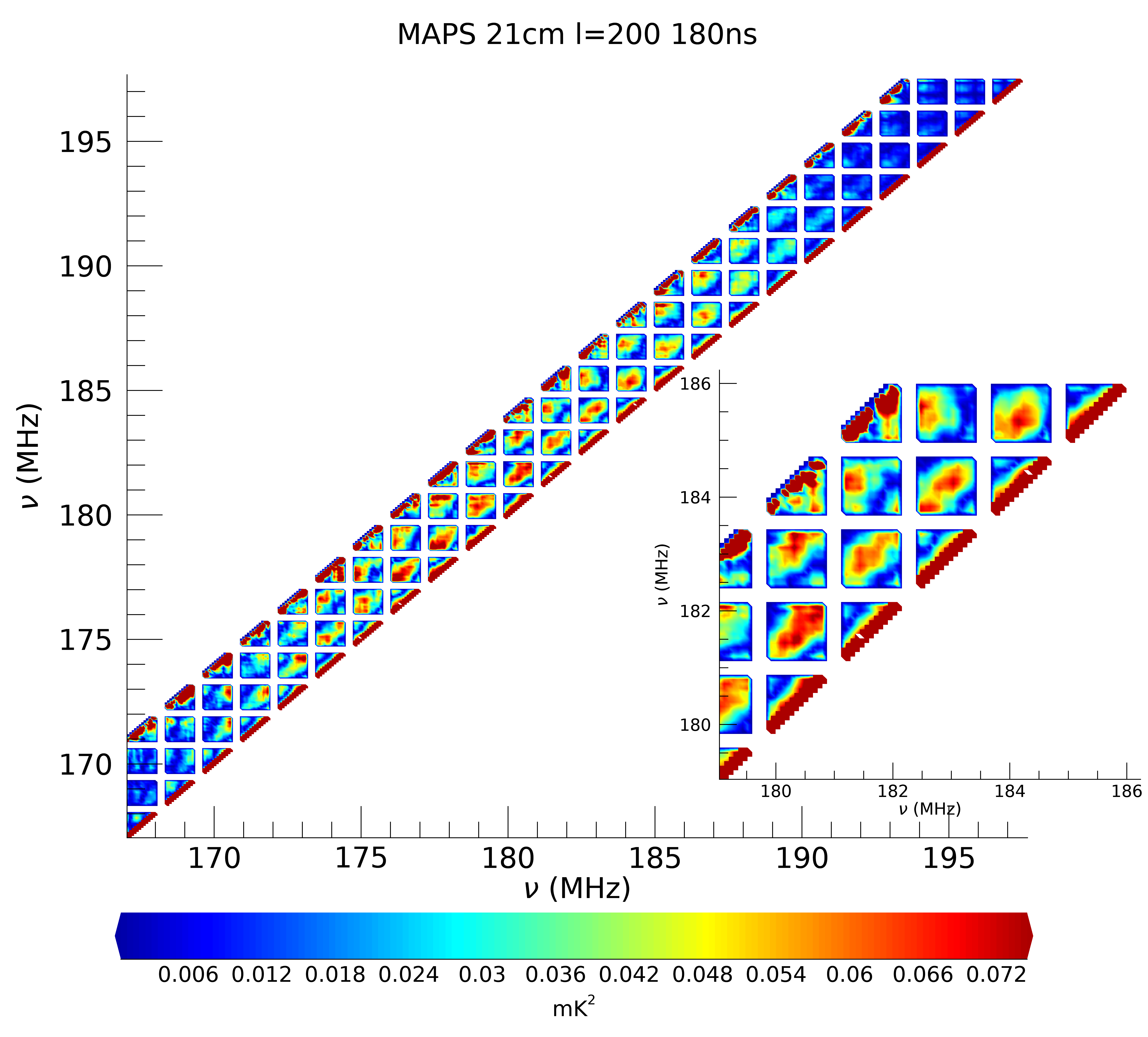}
\includegraphics[width=0.48\textwidth]{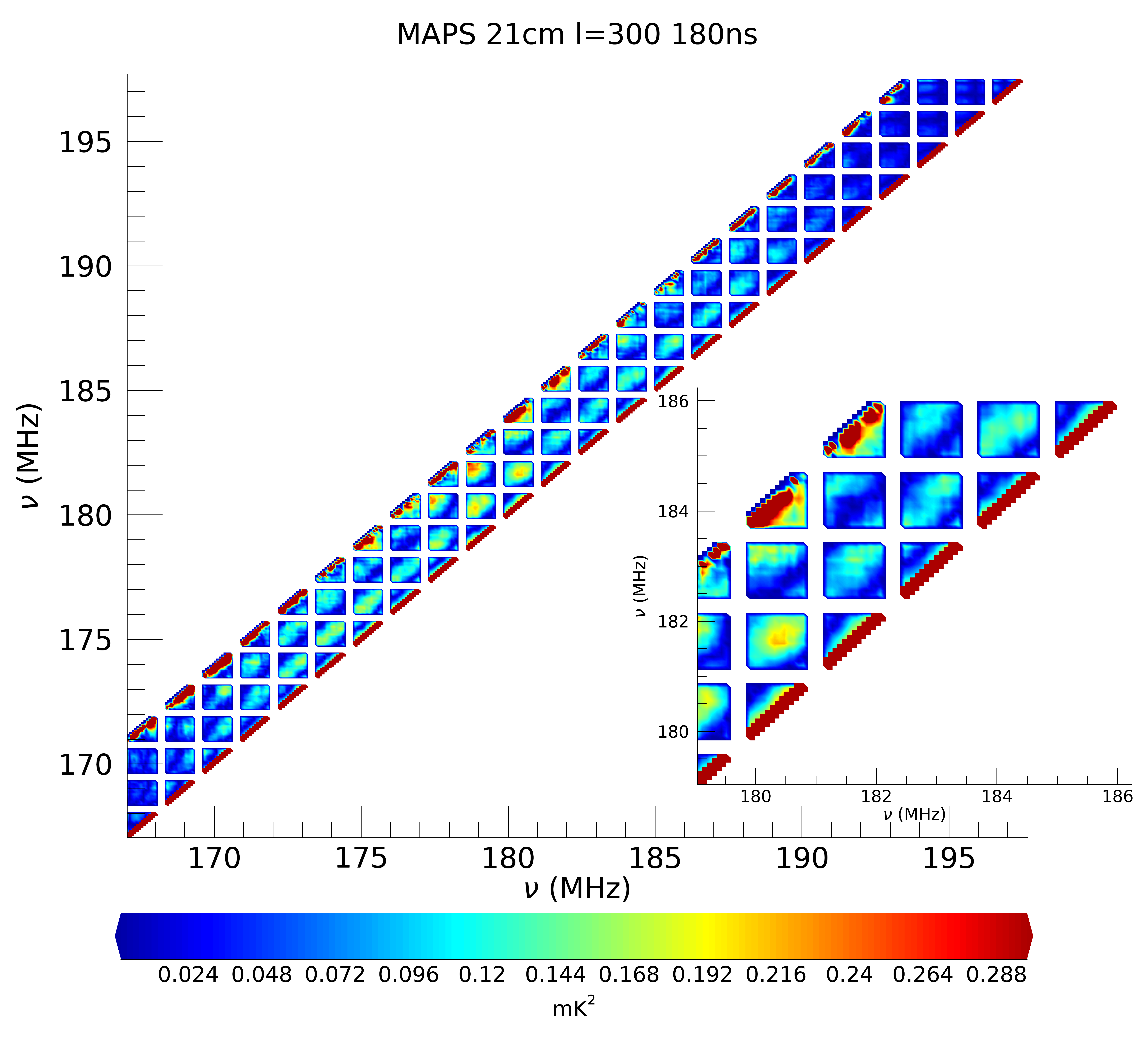}
\caption{Contour maps of dimensionless MAPS for the 21cm simulation and $l=$200, 300. Note the different colour scale for each plot.}
\label{fig:2d_21cm}
\end{figure*}

Figure \ref{fig:2d_data} shows contour plots of the dimensionless MAPS as a function of frequency, for $l=200, 300$.
The inset shows a subset of the data for clarity.
The data have been corrected for the signal loss due to the 180~ns filter for $\Delta\nu <$ 4~MHz. The red diagonal stripe at a spectral difference of $\sim$4~MHz is an artifact of the filter (where the cross power is transitioning from positive to negative, and the ratio of the power to the weights is undefined). Spectral separations larger than 4~MHz have been omitted. In general, there are frequencies (redshifts) where the MAPS is enhanced. Figure \ref{fig:2d_21cm} then shows the equivalent filtered MAPS for the 21~cm simulation. The same filter artifact can be observed here.

\begin{figure*}
\centering
\includegraphics[width=0.48\textwidth]{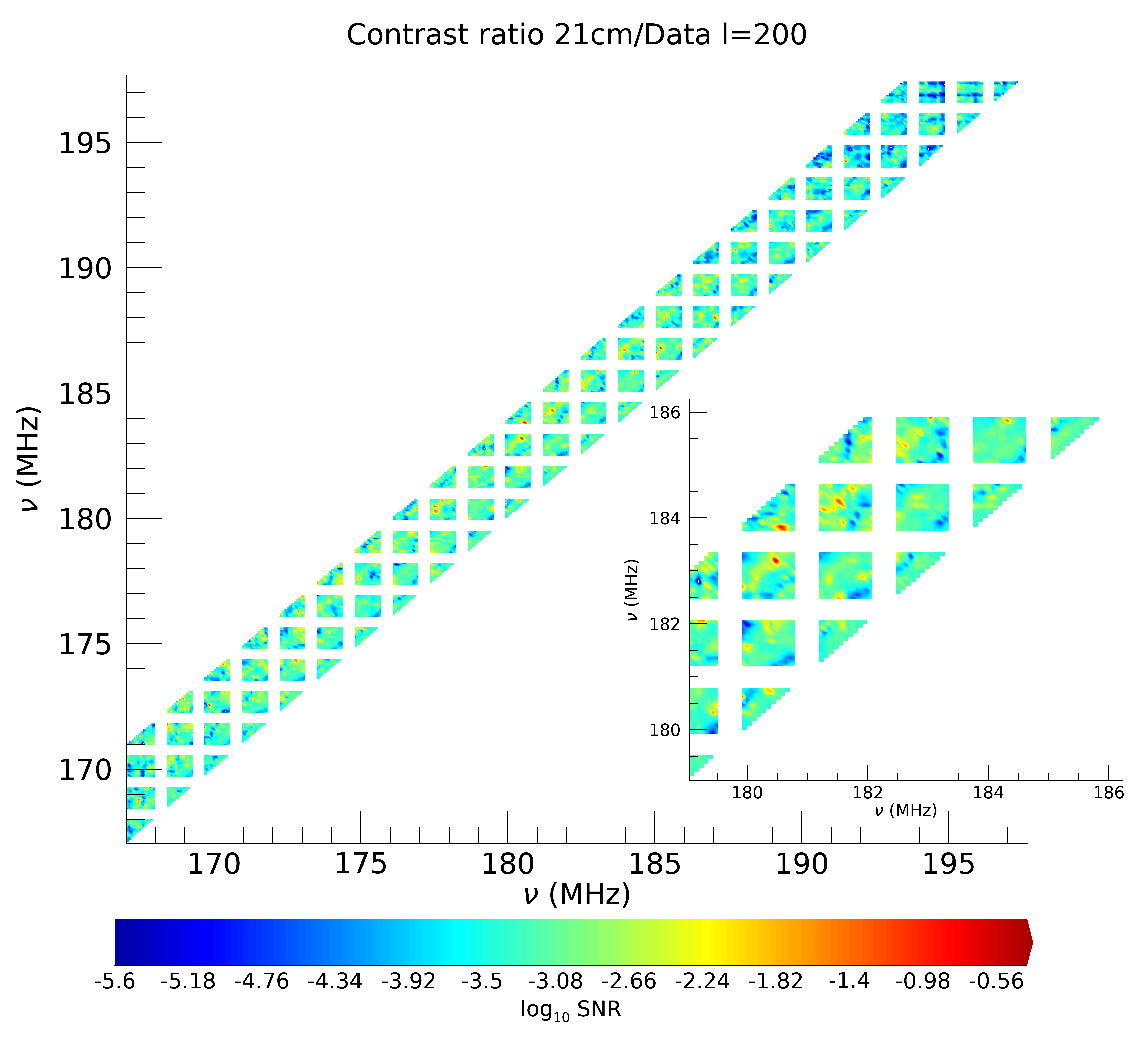}
\includegraphics[width=0.48\textwidth]{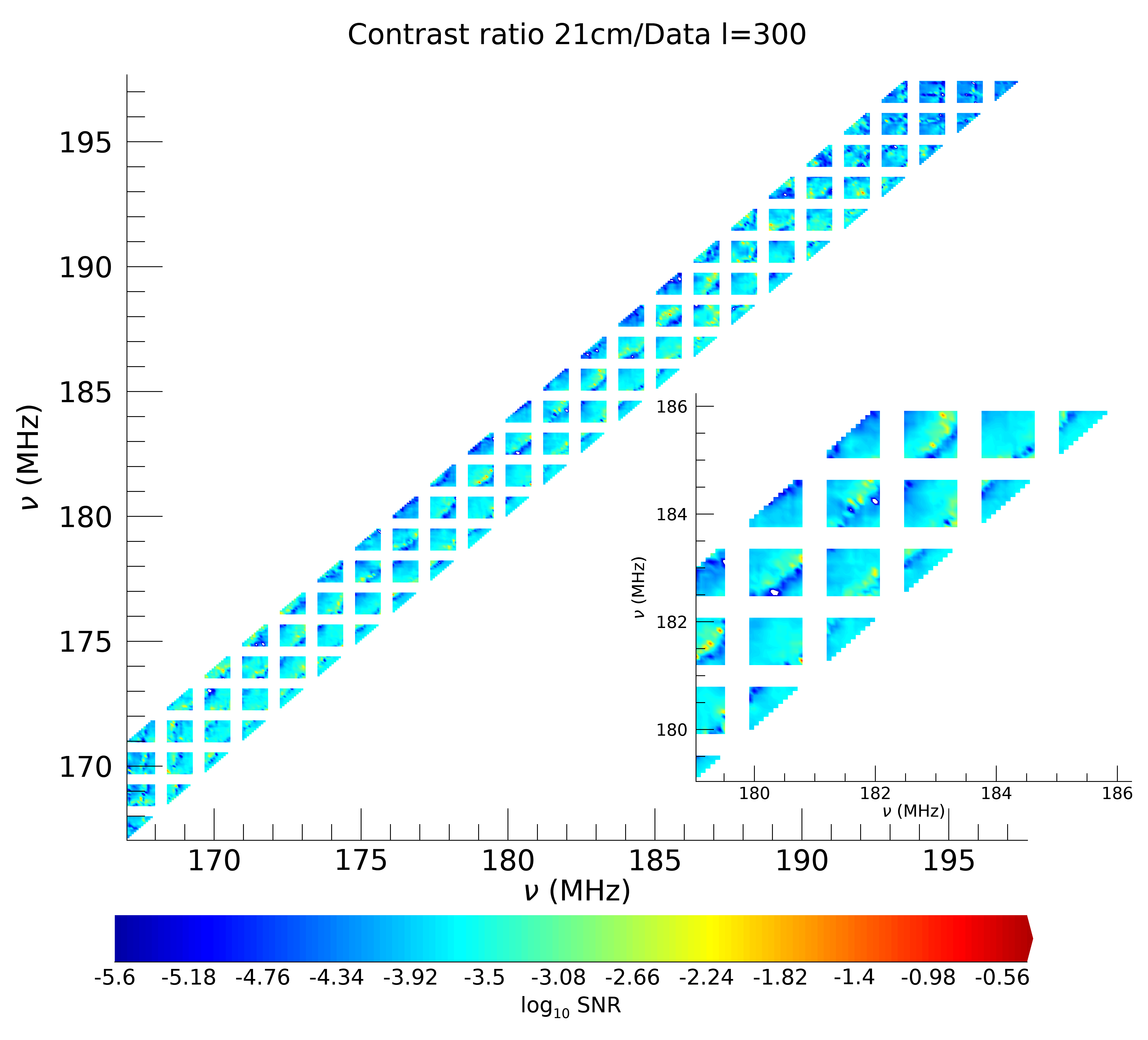}
\caption{Contour maps of contrast ratio of the 21cm power to the data, for $l=$200, 300. The logarithm is plotted, showing a 10$^2$--10$^3$ ratio of data to the expected signal power.}
\label{fig:2d_contrast}
\end{figure*}

Figure \ref{fig:2d_contrast} then shows the contrast ratio between the expected 21~cm power and the measured data (logarithmic scale). The scale is kept the same for both $l$-modes to show the differences.
Greatest contrast is seen for $l=200$ ($k_\bot=0.034h$~Mpc~$^{-1}$) where the contrast ratio is generally 10$^2$--10$^3$, but observed as low as 10$^{1}$ in some regions. The contrast is poorer than for the ergodic MAPS, where data averaging reduces the noise and systematics further. We also tried averaging the data to 160~kHz resolution, and performing the local MAPS on those data. As expected, the contrast between 21cm signal and the data did not change significantly, due to the data being systematics-limited.

\section{Angular Power Spectrum}
The data are also used to compute the single frequency angular power spectrum, to connect with more traditional studies of this statistic, and corresponding to the $\nu_1=\nu_2$ term of Equation \ref{eqn:dcl} and Equation \ref{eqn:aps}. All spectral channels are foreground dominated, and so the same filter is applied to reduce the level of contamination. The single channel case also allows for the data distribution to be studied, in contrast to the sample variance estimator, which uses all of the data blindly in the summation described in Equation \ref{eqn:dcl}.

\begin{figure*}
\centering
\includegraphics[width=0.85\textwidth]{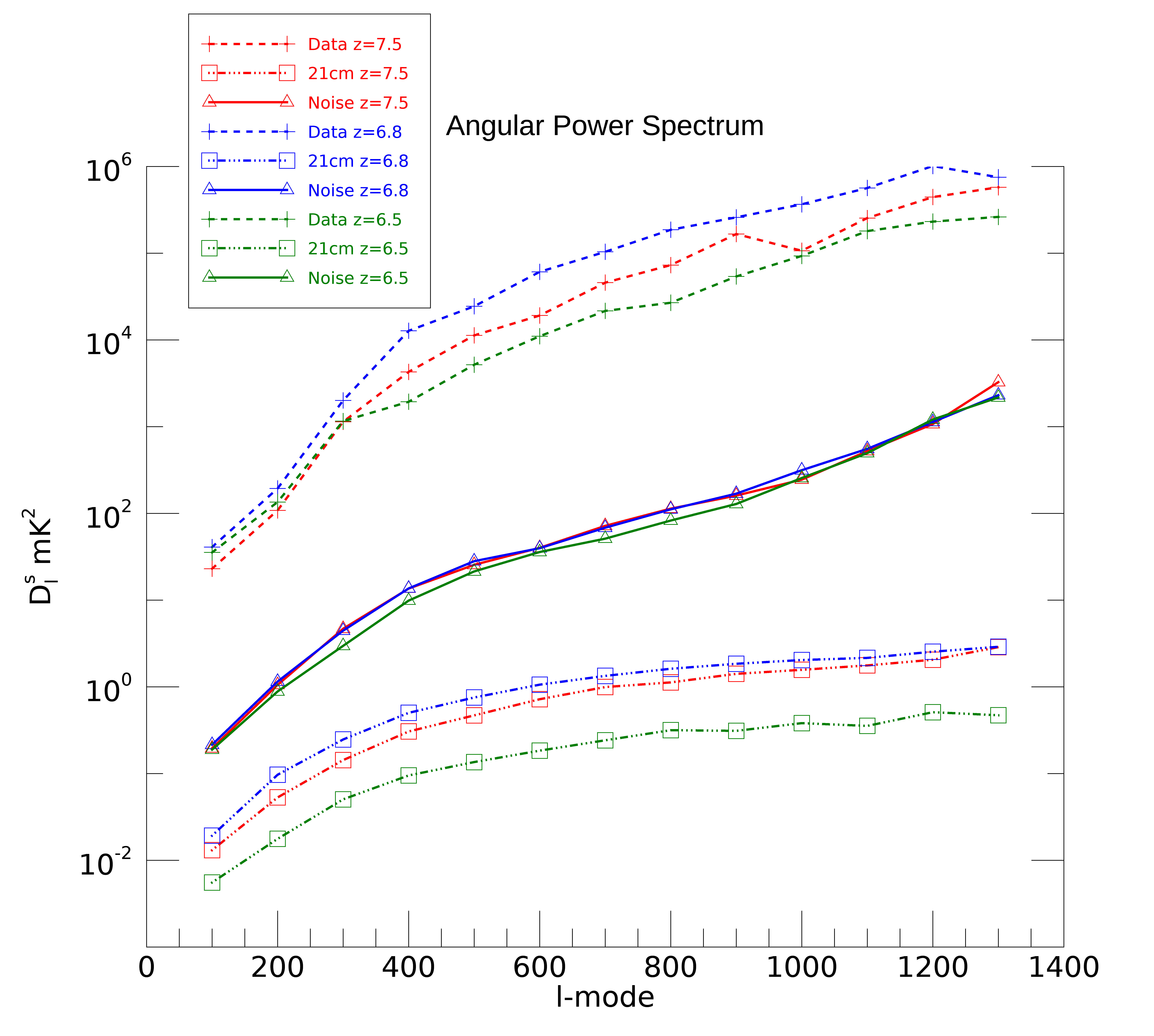}
\caption{Angular power spectra, $D_l^s$ (dashed), 21~cm signal power (dot-dashed) and estimated noise (solid) for three redshifts, $z=6.5, 6.8, 7.5$. Data have been averaged over the central 880~kHz in each 1.28~MHz coarse channel.}
\label{fig:aps}
\end{figure*}

Figure \ref{fig:aps} shows data, 21~cm and noise dimensionless angular power spectra as a function of $l$ and for a set of redshifts, after application of the same 180~ns filter. The data are binned into coarse $\Delta{l}=100$ ($\Delta k_\bot=0.016h$~Mpc~$^{-1}$) bins, and have been averaged over the central 880~kHz in each 1.28~MHz coarse channel.
The filtered data show the lowest values at low redshift, but are significantly higher than the expected signal with a contrast that exceeds that observed in the spherically-averaged power spectrum, but only by a factor of 3--5. The expected 21~cm signal is also lowest at this redshift. The residual foreground power in each spectral channel is therefore less easily treated with the filter than with a line-of-sight spectral transform, but the results are not too degraded compared with the spherically-averaged power spectrum, and allow for greater granularity to study the signal evolution. 

The variance (power spectrum) can also be extracted from the data histograms. Typically, the data would be expected to be close to Gaussian-distributed, reflecting the near-Gaussian expectation for the signal, and the Gaussian radiometric noise. Estimation of the data variance directly from the sample data histograms should yield the same value as the sample variance computation (which is used for power spectrum estimation normally, e.g. Equation \ref{eqn:dcl}). Presence of bright foregrounds may skew this distribution, leading to outlier $k$-modes in the dataset.

\begin{figure*}
\centering
\includegraphics[width=0.48\textwidth]{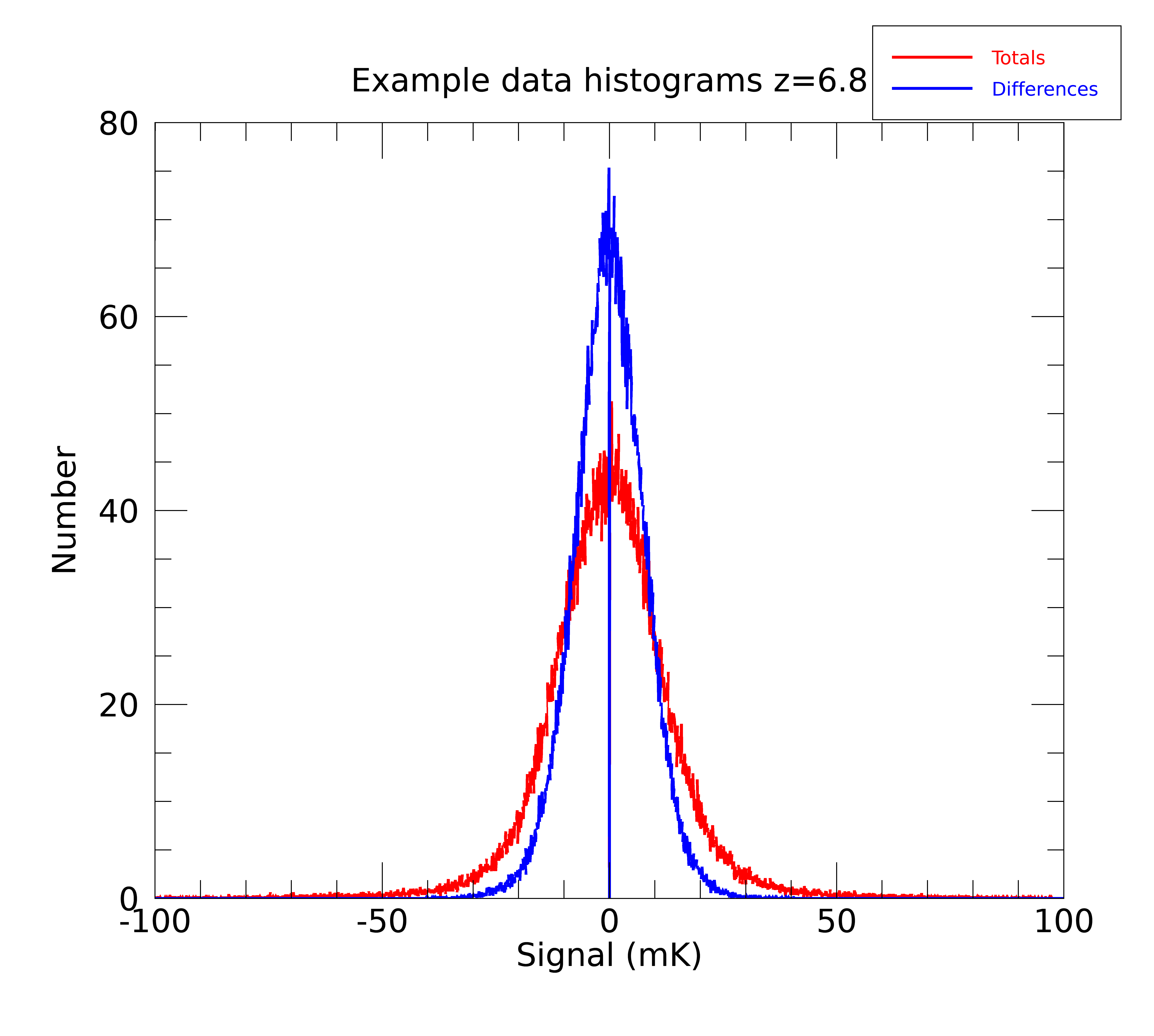}
\includegraphics[width=0.48\textwidth]{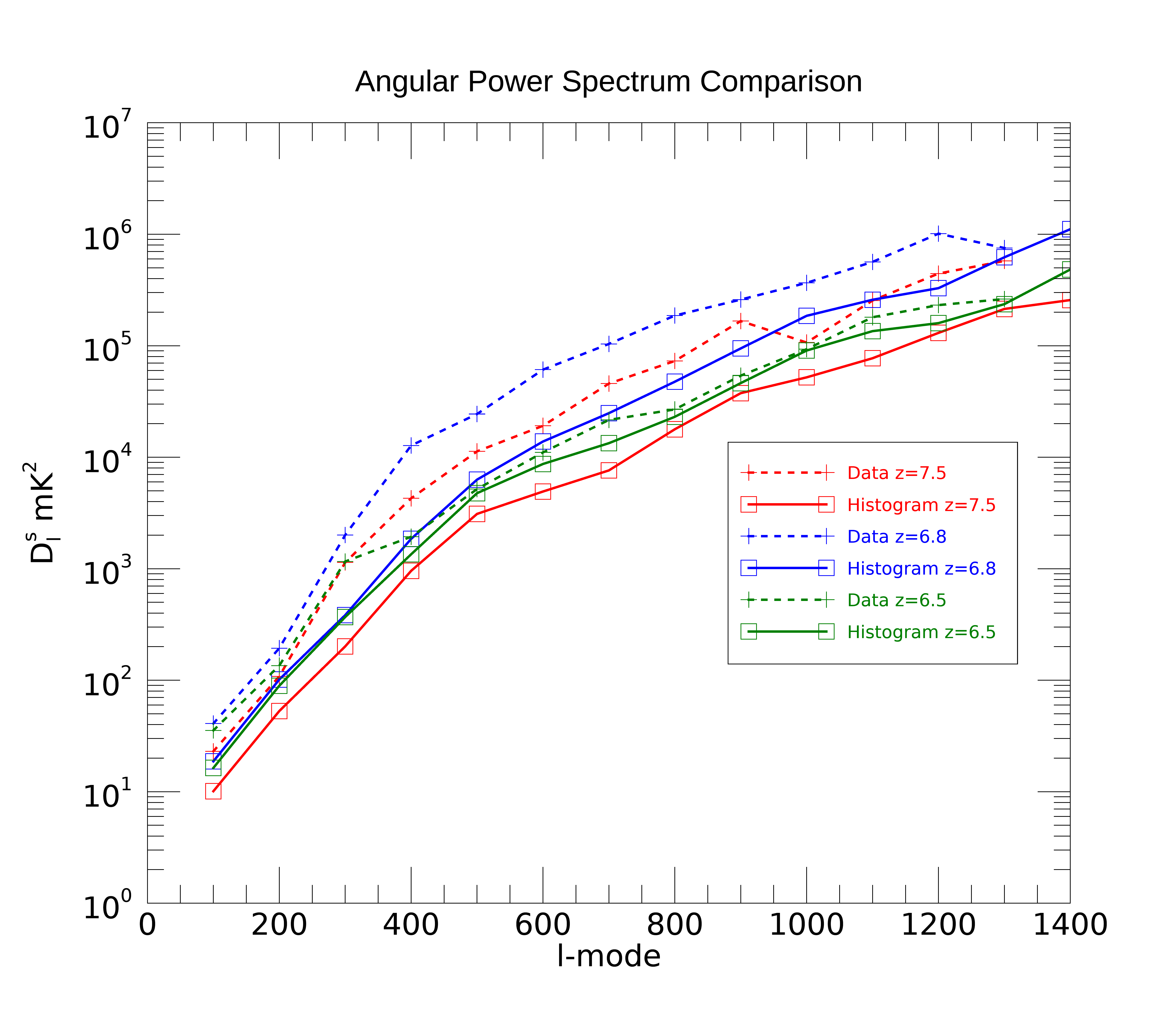}
\caption{(Left) Histograms of the real and imaginary components of the gridded visibilities for a 880~kHz band at $z=6.8$ and $l=200$. (Right) Angular power spectra across three redshifts using the normal sample variance estimate (dashed) and extracted from fits to the histograms (solid).}
\label{fig:apsh}
\end{figure*}

To compute the sample data distributions, the real and imaginary components of the gridded visibilities are used to compute a histogram of the data for each $l$-mode, and at each coarse channel, with data being combined across spectral channels to form an effective 880~kHz bandwidth. A Gaussian is then fitted to the gridded visibilities of the `totals' and the `differences' to estimate the sample variance. A histogram resolution of 0.1 mK is chosen to avoid discretization effects. Figure~\ref{fig:apsh} (left) shows example histogrammed data for the dimensionless angular power spectrum  for totals visibilities (red) and the differenced visibilities (blue), for $l=200$ and $z=6.8$. Figure \ref{fig:apsh} (right) shows the comparison of the angular power spectra using the sample variance (dashed) and computed from Gaussian fits to the histograms (solid).
Use of the histograms generally yields improved results (lower power) across redshift and angular mode. At $l=200$, where some of the best results have been found with these data, the angular power is 53~mK$^2$ at $z=7.5$, compared with the 21~cm simulated power of 0.05~mK$^2$, yielding a data-to-signal contrast ratio of $\sim$1000, which exceeds that found in \citet{trott20} for the spherically-averaged power spectrum with the same data, by a factor of 3--5. A similar ratio is obtained at $l=100$, where the measured power is $\Delta^2=10$~mK$^2$.

\section{Discussion}
The foreground filter does a reasonable job of removing a lot of the correlated spectral structure, but there remains spectral structure on EoR scales that leaves excess power in the MAPS. In spherically-averaged power spectrum estimation, a line-of-sight spectral transform must be performed to move from configuration to $k$-space, and this is problematic due to the data being discrete and band-limited. Typically, a spectral taper is applied prior to transform to suppress sidelobes, but this is prone to error when applied to discrete data, and comes at the expense of a broader main lobe (correlating line-of-sight $k$-modes). Several groups have studied the effect of the taper application and spectral transform on leakage of foreground power into EoR modes \citep[e.g.,][]{lanman20,barry22}, showing that leaked power can be a source of excess power. With MAPS, no such transform is required, and foregrounds need to be treated separately. Thus, the residual (excess) power relative to the noise level observed in these data is indicative of inherent spectral variability on those scales, and not from leaked power from larger modes. This conclusion is not unreasonable, because the data calibration only considered unpolarized point sources and extended foregrounds, with unpolarized diffuse and polarized diffuse, and point source models omitted. Polarized emission, which is known to be significant in this observing field \citep{bernardi13,lenc17} will imprint spectral structure for sources with non-zero Faraday depth.

In general, the single-frequency angular power spectrum yields a higher contrast ratio than the local MAPS, suggesting that there are benefits to considering small spectral differences, where the 21~cm signal power is reduced, but the foregrounds also partially decohere.

\section{Conclusions}
The Multi-Frequency Angular Power Spectrum (MAPS) has been applied to a deep 110-hour integration of the EoR0 field from the MWA EoR project in $z=6.5-7.5$. These data have been previously used with a spherically-averaged power spectrum to produce scientifically-relevant upper limits on the power spectrum of brightness temperature fluctuations in the hyperfine transition line of cosmological neutral hydrogen. The angular power spectrum has the advantage of not requiring a band-limited line-of-sight spectral transform, which mixes line-of-sight $k$-modes and needs to be performed on (approximately) ergodic subsets of a full dataset. However, it suffers from contamination from residual continuum foreground signal, which is highly-dominant and cannot easily be distinguished without spectral information. Here, a filter is applied to the broadband dataset prior to estimation of the angular power spectrum, to remove smoothly-varying signal structure. This improves the power spectrum estimation by two orders-of-magnitude, but still yields poorer results relative to the expected 21~cm signal compared with the spherically-averaged power spectrum.

Treatment of foregrounds differs between the angular and spherically-averaged power spectrum approaches; the former using a filter with the latter using the properties of the Fourier Transform. The filter employed here is shown to improve its performance with larger initial bandwidths \citep{dayenu}, whereas the Fourier Transform cannot be used over bandwidths that destroy the assumption of signal ergodicity. Thus, increasing the bandwidth of experimental data may improve the performance of the APS compared with the spherically-averaged power spectrum. Additionally, other foreground filters may be explored and employed \citep{pal21}.
In future, a combination of spectral and spatial \citep[e.g., GMCA, ][]{chapman14} filtering may be required to yield improved results.

\begin{acknowledgement}
This research was partly supported by the Australian Research Council Centre of Excellence for All Sky Astrophysics in 3 Dimensions (ASTRO 3D), through project number CE170100013. CMT is supported by an ARC Future Fellowship under grant FT180100321.
The International Centre for Radio Astronomy Research (ICRAR) is a Joint Venture of Curtin University and The University of Western Australia, funded by the Western Australian State government.
The MWA Phase II upgrade project was supported by Australian Research Council LIEF grant LE160100031 and the Dunlap Institute for Astronomy and Astrophysics at the University of Toronto.
This scientific work makes use of the Murchison Radio-astronomy Observatory, operated by CSIRO. We acknowledge the Wajarri Yamatji people as the traditional owners of the Observatory site. Support for the operation of the MWA is provided by the Australian Government (NCRIS), under a contract to Curtin University administered by Astronomy Australia Limited. We acknowledge the Pawsey Supercomputing Centre which is supported by the Western Australian and Australian Governments. Data were processed at the Pawsey Supercomputing Centre. RM is supported by the Wenner-Gren Postdoctoral Fellowship and GM acknowledges support by Swedish Research Council grant 2020-04691. RM is also supported by the Israel Academy of Sciences and Humanities \& Council for Higher Education Excellence Fellowship Program for International Postdoctoral Researchers.
\end{acknowledgement}

%
%

\bibliographystyle{aa} 

\begin{thebibliography}{31}
\expandafter\ifx\csname natexlab\endcsname\relax\def\natexlab#1{#1}\fi

\bibitem[{{Barry} {et~al.}(2019){Barry}, {Beardsley}, {Byrne}, {Hazelton},
  {Morales}, {Pober}, \& {Sullivan}}]{barry19}
{Barry}, N., {Beardsley}, A.~P., {Byrne}, R., {et~al.} 2019, arXiv e-prints
  1901.02980 [\eprint[arXiv]{1901.02980}]

\bibitem[{{Barry} \& {Chokshi}(2022)}]{barry22}
{Barry}, N. \& {Chokshi}, A. 2022, \apj, 929, 64

\bibitem[{{Bernardi} {et~al.}(2013){Bernardi}, {Greenhill}, {Mitchell}, {Ord},
  {Hazelton}, {Gaensler}, {de Oliveira-Costa}, {Morales}, {Udaya Shankar},
  {Subrahmanyan}, {Wayth}, {Lenc}, {Williams}, {Arcus}, {Arora}, {Barnes},
  {Bowman}, {Briggs}, {Bunton}, {Cappallo}, {Corey}, {Deshpande}, {deSouza},
  {Emrich}, {Goeke}, {Herne}, {Hewitt}, {Johnston-Hollitt}, {Kaplan}, {Kasper},
  {Kincaid}, {Koenig}, {Kratzenberg}, {Lonsdale}, {Lynch}, {McWhirter},
  {Morgan}, {Oberoi}, {Pathikulangara}, {Prabu}, {Remillard}, {Rogers},
  {Roshi}, {Salah}, {Sault}, {Srivani}, {Stevens}, {Tingay}, {Waterson},
  {Webster}, {Whitney}, {Williams}, \& {Wyithe}}]{bernardi13}
{Bernardi}, G., {Greenhill}, L.~J., {Mitchell}, D.~A., {et~al.} 2013, \apj,
  771, 105

\bibitem[{{Bharadwaj} \& {Ali}(2005)}]{2005MNRAS.356.1519B}
{Bharadwaj}, S. \& {Ali}, S.~S. 2005, \mnras, 356, 1519

\bibitem[{{Bowman} {et~al.}(2013){Bowman}, {Cairns}, {Kaplan}, {Murphy},
  {Oberoi}, \& {others}}]{bowman13_mwascience}
{Bowman}, J.~D., {Cairns}, I., {Kaplan}, D.~L., {et~al.} 2013, PASA, 30, 31

\bibitem[{{Bowman} {et~al.}(2018){Bowman}, {Rogers}, {Monsalve}, {Mozdzen}, \&
  {Mahesh}}]{edges18}
{Bowman}, J.~D., {Rogers}, A. E.~E., {Monsalve}, R.~A., {Mozdzen}, T.~J., \&
  {Mahesh}, N. 2018, \nat, 555, 67

\bibitem[{{Chapman} {et~al.}(2014){Chapman}, {Zaroubi}, \&
  {Abdalla}}]{chapman14}
{Chapman}, E., {Zaroubi}, S., \& {Abdalla}, F. 2014, ArXiv/1408.4695
  [\eprint[arXiv]{1408.4695}]

\bibitem[{{Datta} {et~al.}(2007){Datta}, {Choudhury}, \&
  {Bharadwaj}}]{2007MNRAS.378..119D}
{Datta}, K.~K., {Choudhury}, T.~R., \& {Bharadwaj}, S. 2007, \mnras, 378, 119

\bibitem[{{DeBoer} {et~al.}(2017){DeBoer}, {Parsons}, {Aguirre}, {Alexander},
  {Ali}, {Beardsley}, {Bernardi}, {Bowman}, {Bradley}, {Carilli}, {Cheng}, {de
  Lera Acedo}, {Dillon}, {Ewall-Wice}, {Fadana}, {Fagnoni}, {Fritz},
  {Furlanetto}, {Glendenning}, {Greig}, {Grobbelaar}, {Hazelton}, {Hewitt},
  {Hickish}, {Jacobs}, {Julius}, {Kariseb}, {Kohn}, {Lekalake}, {Liu}, {Loots},
  {MacMahon}, {Malan}, {Malgas}, {Maree}, {Martinot}, {Mathison}, {Matsetela},
  {Mesinger}, {Morales}, {Neben}, {Patra}, {Pieterse}, {Pober}, {Razavi-Ghods},
  {Ringuette}, {Robnett}, {Rosie}, {Sell}, {Smith}, {Syce}, {Tegmark},
  {Thyagarajan}, {Williams}, \& {Zheng}}]{deboer17}
{DeBoer}, D.~R., {Parsons}, A.~R., {Aguirre}, J.~E., {et~al.} 2017, \pasp, 129,
  045001

\bibitem[{{Ewall-Wice} {et~al.}(2021){Ewall-Wice}, {Kern}, {Dillon}, {Liu},
  {Parsons}, {Singh}, {Lanman}, {La Plante}, {Fagnoni}, {Acedo}, {DeBoer},
  {Nunhokee}, {Bull}, {Chang}, {Lazio}, {Aguirre}, \& {Weinberg}}]{dayenu}
{Ewall-Wice}, A., {Kern}, N., {Dillon}, J.~S., {et~al.} 2021, \mnras, 500, 5195

\bibitem[{{Greig} {et~al.}(2022){Greig}, {Wyithe}, {Murray}, {Mutch}, \&
  {Trott}}]{greig22}
{Greig}, B., {Wyithe}, J., {Murray}, S.~G., {Mutch}, S.~J., \& {Trott}, C.~M.
  2022, \mnras

\bibitem[{{Koopmans} {et~al.}(2015){Koopmans}, {Pritchard}, {Mellema},
  {Aguirre}, {Ahn}, {Barkana}, {van Bemmel}, {Bernardi}, {Bonaldi}, {Briggs},
  {de Bruyn}, {Chang}, {Chapman}, {Chen}, {Ciardi}, {Dayal}, {Ferrara},
  {Fialkov}, {Fiore}, {Ichiki}, {Illiev}, {Inoue}, {Jelic}, {Jones}, {Lazio},
  {Maio}, {Majumdar}, {Mack}, {Mesinger}, {Morales}, {Parsons}, {Pen},
  {Santos}, {Schneider}, {Semelin}, {de Souza}, {Subrahmanyan}, {Takeuchi},
  {Vedantham}, {Wagg}, {Webster}, {Wyithe}, {Datta}, \& {Trott}}]{koopmans15}
{Koopmans}, L., {Pritchard}, J., {Mellema}, G., {et~al.} 2015, Advancing
  Astrophysics with the Square Kilometre Array (AASKA14), 1

\bibitem[{{Lanman} {et~al.}(2020){Lanman}, {Pober}, {Kern}, {de Lera Acedo},
  {DeBoer}, \& {Fagnoni}}]{lanman20}
{Lanman}, A.~E., {Pober}, J.~C., {Kern}, N.~S., {et~al.} 2020, \mnras, 494,
  3712

\bibitem[{{Lenc} {et~al.}(2017){Lenc}, {Anderson}, {Barry}, {Bowman}, {Cairns},
  {Farnes}, {Gaensler}, {Heald}, {Johnston-Hollitt}, {Kaplan}, {Lynch},
  {McCauley}, {Mitchell}, {Morgan}, {Morales}, {Murphy}, {Offringa}, {Ord},
  {Pindor}, {Riseley}, {Sadler}, {Sobey}, {Sokolowski}, {Sullivan},
  {O'Sullivan}, {Sun}, {Tremblay}, {Trott}, \& {Wayth}}]{lenc17}
{Lenc}, E., {Anderson}, C.~S., {Barry}, N., {et~al.} 2017, \pasa, 34, e040

\bibitem[{Line(2022)}]{Line2022}
Line, J. L.~B. 2022, Journal of Open Source Software, 7, 3676

\bibitem[{{Mesinger} {et~al.}(2011){Mesinger}, {Furlanetto}, \&
  {Cen}}]{mesinger11}
{Mesinger}, A., {Furlanetto}, S., \& {Cen}, R. 2011, \mnras, 411, 955

\bibitem[{{Mondal} {et~al.}(2018){Mondal}, {Bharadwaj}, \& {Datta}}]{mondal18}
{Mondal}, R., {Bharadwaj}, S., \& {Datta}, K.~K. 2018, \mnras, 474, 1390

\bibitem[{{Mondal} {et~al.}(2019){Mondal}, {Bharadwaj}, {Iliev}, {Datta},
  {Majumdar}, {Shaw}, \& {Sarkar}}]{mondal19}
{Mondal}, R., {Bharadwaj}, S., {Iliev}, I.~T., {et~al.} 2019, \mnras, 483, L109

\bibitem[{{Mondal} {et~al.}(2022){Mondal}, {Mellema}, {Murray}, \&
  {Greig}}]{mondal22}
{Mondal}, R., {Mellema}, G., {Murray}, S.~G., \& {Greig}, B. 2022, arXiv
  e-prints, arXiv:2203.11095

\bibitem[{{Mondal} {et~al.}(2020){Mondal}, {Shaw}, {Iliev}, {Bharadwaj},
  {Datta}, {Majumdar}, {Sarkar}, \& {Dixon}}]{mondal20}
{Mondal}, R., {Shaw}, A.~K., {Iliev}, I.~T., {et~al.} 2020, \mnras, 494, 4043

\bibitem[{{Pal} {et~al.}(2021){Pal}, {Bharadwaj}, {Ghosh}, \&
  {Choudhuri}}]{pal21}
{Pal}, S., {Bharadwaj}, S., {Ghosh}, A., \& {Choudhuri}, S. 2021, \mnras, 501,
  3378

\bibitem[{{Park} {et~al.}(2019){Park}, {Mesinger}, {Greig}, \&
  {Gillet}}]{park19}
{Park}, J., {Mesinger}, A., {Greig}, B., \& {Gillet}, N. 2019, \mnras, 484, 933

\bibitem[{{Parsons} \& {Backer}(2009)}]{parsons09}
{Parsons}, A.~R. \& {Backer}, D.~C. 2009, AJ, 138, 219

\bibitem[{{Santos} {et~al.}(2005){Santos}, {Cooray}, \&
  {Knox}}]{2005ApJ...625..575S}
{Santos}, M.~G., {Cooray}, A., \& {Knox}, L. 2005, \apj, 625, 575

\bibitem[{{Singh} {et~al.}(2022){Singh}, {Nambissan T.}, {Subrahmanyan}, {Udaya
  Shankar}, {Girish}, {Raghunathan}, {Somashekar}, {Srivani}, \&
  {Sathyanarayana Rao}}]{saras322}
{Singh}, S., {Nambissan T.}, J., {Subrahmanyan}, R., {et~al.} 2022, Nature
  Astronomy [\eprint[arXiv]{2112.06778}]

\bibitem[{{Thekkeppattu} {et~al.}(2022){Thekkeppattu}, {McKinley}, {Trott},
  {Jones}, \& {Ung}}]{2022arXiv220310466T}
{Thekkeppattu}, J.~N., {McKinley}, B., {Trott}, C.~M., {Jones}, J., \& {Ung},
  D. C.~X. 2022, arXiv e-prints, arXiv:2203.10466

\bibitem[{{Tingay} {et~al.}(2013){Tingay}, {Goeke}, {Bowman}, {Emrich}, \&
  {others}}]{tingay13_mwasystem}
{Tingay}, S.~J., {Goeke}, R., {Bowman}, J.~D., {Emrich}, D., \& {others}. 2013,
  PASA, 30, 7

\bibitem[{{Trott} {et~al.}(2020){Trott}, {Jordan}, {Midgley}, {Barry}, {Greig},
  {Pindor}, {Cook}, {Sleap}, {Tingay}, {Ung}, {Hancock}, {Williams}, {Bowman},
  {Byrne}, {Chokshi}, {Hazelton}, {Hasegawa}, {Jacobs}, {Joseph}, {Li}, {Line},
  {Lynch}, {McKinley}, {Mitchell}, {Morales}, {Ouchi}, {Pober}, {Rahimi},
  {Takahashi}, {Wayth}, {Webster}, {Wilensky}, {Wyithe}, {Yoshiura}, {Zhang},
  \& {Zheng}}]{trott20}
{Trott}, C.~M., {Jordan}, C.~H., {Midgley}, S., {et~al.} 2020, \mnras, 493,
  4711

\bibitem[{{Trott} {et~al.}(2016){Trott}, {Pindor}, {Procopio}, {Wayth},
  {Mitchell}, {McKinley}, {Tingay}, {Barry}, {Beardsley}, {Bernardi}, {Bowman},
  {Briggs}, {Cappallo}, {Carroll}, {de Oliveira-Costa}, {Dillon}, {Ewall-Wice},
  {Feng}, {Greenhill}, {Hazelton}, {Hewitt}, {Hurley-Walker},
  {Johnston-Hollitt}, {Jacobs}, {Kaplan}, {Kim}, {Lenc}, {Line}, {Loeb},
  {Lonsdale}, {Morales}, {Morgan}, {Neben}, {Thyagarajan}, {Oberoi},
  {Offringa}, {Ord}, {Paul}, {Pober}, {Prabu}, {Riding}, {Udaya Shankar},
  {Sethi}, {Srivani}, {Subrahmanyan}, {Sullivan}, {Tegmark}, {Webster},
  {Williams}, {Williams}, {Wu}, \& {Wyithe}}]{trott16chips}
{Trott}, C.~M., {Pindor}, B., {Procopio}, P., {et~al.} 2016, \apj, 818, 139

\bibitem[{{van Haarlem} {et~al.}(2013){van Haarlem}, {Wise}, {Gunst}, {Heald},
  {McKean}, {Hessels}, {de Bruyn}, {Nijboer}, {Swinbank}, {Fallows},
  {Brentjens}, {Nelles}, {Beck}, {Falcke}, {Fender}, {H{\"o}randel},
  {Koopmans}, {Mann}, {Miley}, {R{\"o}ttgering}, {Stappers}, \&
  {others}}]{vanhaarlem13}
{van Haarlem}, M.~P., {Wise}, M.~W., {Gunst}, A.~W., {et~al.} 2013, \aap, 556,
  A2

\bibitem[{{Wayth} {et~al.}(2018){Wayth}, {Tingay}, {Trott}, {Emrich},
  {Johnston-Hollitt}, {McKinley}, {Gaensler}, {Beardsley}, {Booler}, {Crosse},
  {Franzen}, {Horsley}, {Kaplan}, {Kenney}, {Morales}, {Pallot}, {Sleap},
  {Steele}, {Walker}, {Williams}, {Wu}, {Cairns}, {Filipovic}, {Johnston},
  {Murphy}, {Quinn}, {Staveley-Smith}, {Webster}, \& {Wyithe}}]{wayth18}
{Wayth}, R.~B., {Tingay}, S.~J., {Trott}, C.~M., {et~al.} 2018, \pasa, 35
  [\eprint[arXiv]{1809.06466}]

\end{thebibliography}

\end{document}